\documentclass[sigconf]{acmart}

\usepackage{hyperref}
\usepackage{amsmath,amsfonts}
\usepackage{graphicx}
\usepackage{textcomp}
\usepackage{xcolor}
\usepackage{mathtools}
\usepackage{multirow}
\usepackage{array}
\usepackage{bm}
\usepackage{booktabs} 
\usepackage{xspace}
\usepackage{enumitem}
\usepackage{subcaption}

\def\BibTeX{{\rm B\kern-.05em{\sc i\kern-.025em b}\kern-.08em
    T\kern-.1667em\lower.7ex\hbox{E}\kern-.125emX}}

\copyrightyear{2025}
\acmYear{2025}
\setcopyright{rightsretained}
\acmConference[GLSVLSI '25]{Proceedings of the Great Lakes Symposium on VLSI 2025}{June 30--July 2, 2025}{New Orleans, LA, USA}
\acmBooktitle{Proceedings of the Great Lakes Symposium on VLSI 2025 (GLSVLSI '25), June 30--July 2, 2025, New Orleans, LA, USA}
\acmDOI{10.1145/3716368.3735241}

\makeatletter
\gdef\@copyrightpermission{
  \begin{minipage}{0.3\columnwidth}
   \href{https://creativecommons.org/licenses/by-nc-nd/4.0/}{\includegraphics[width=0.9\columnwidth]{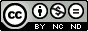}}
  \end{minipage}\hfill
  \begin{minipage}{0.7\columnwidth}
   This work is under a \href{https://creativecommons.org/licenses/by-nc-nd/4.0/}{Creative Commons Attribution-NonCommercial-NoDerivs International 4.0 License.}
  \end{minipage}
  \vspace{5pt}
}

\makeatother

\begin{document}

\title{
Robust Reasoning and Learning with Brain-Inspired Representations under Hardware-Induced Nonlinearities}

\author{William Youngwoo Chung}
\email{chungwy1@uci.edu}
\orcid{0000-0002-0500-4647}
\affiliation{%
  \institution{University of California, Irvine}
  \city{Irvine}
  \state{California}
  \country{USA}
}
\author{Hamza Errahmouni Barkam}
\email{herrahmo@uci.edu}
\affiliation{%
  \institution{University of California, Irvine}
  \city{Irvine}
  \state{California}
  \country{USA}
}

\author{Tamoghno Das}
\affiliation{%
  \institution{University of California, Irvine}
  \city{Irvine}
  \state{California}
  \country{USA}
}

\author{Mohsen Imani}
\email{m.imani@uci.edu}
\affiliation{%
  \institution{University of California, Irvine}
  \city{Irvine}
  \state{California}
  \country{USA}
}

\begin{abstract}
Traditional machine learning depends on high-precision arithmetic and near-ideal hardware assumptions, which is increasingly challenged by variability in aggressively scaled semiconductor devices. Compute-in-memory (CIM) architectures alleviate data-movement bottlenecks and improve energy efficiency yet introduce nonlinear distortions and reliability concerns. We address these issues with a hardware-aware optimization framework based on Hyperdimensional Computing (HDC), systematically compensating for non-ideal similarity computations in CIM. Our approach formulates encoding as an optimization problem, minimizing the Frobenius norm between an ideal kernel and its hardware-constrained counterpart, and employs a joint optimization strategy for end-to-end calibration of hypervector representations. Experimental results demonstrate that our method when applied to QuantHD achieves 84\% accuracy under severe hardware-induced perturbations, a 48\% increase over naive QuantHD under the same conditions. Additionally, our optimization is vital for graph-based HDC reliant on precise variable-binding for interpretable reasoning. Our framework preserves the accuracy of RelHD on the Cora dataset, achieving a 5.4$\times$ accuracy improvement over naive RelHD under nonlinear environments. By preserving HDC’s robustness and symbolic properties, our solution enables scalable, energy-efficient intelligent systems capable of classification and reasoning on emerging CIM hardware.
\end{abstract}

\ccsdesc[500]{Computing methodologies~Machine learning approaches}
\ccsdesc[500]{Computing methodologies~Symbolic and algebraic manipulation}
\ccsdesc[500]{Hardware~Emerging technologies}

\keywords{Hyperdimensional Computing, Compute-in-Memory, Symbolic Reasoning, Hardware-aware Optimization, Robust Machine Learning, Non-Volatile Memory}

\maketitle
\section{Introduction}
Machine learning (ML), particularly deep learning models, has profoundly influenced numerous domains, delivering exceptional performance in complex tasks such as image recognition, natural language processing, and autonomous decision-making systems. However, these approaches rely heavily on precise, deterministic computation typically executed with high numerical precision (e.g., 16-bit or 32-bit), inherently assuming the reliability and accuracy of the underlying hardware~\cite{hd_efficient}. Historically, the scaling of Complementary Metal-Oxide Semiconductor (CMOS) technology allowed algorithm designers to overlook device-level imperfections, focusing primarily on maximizing computational performance and accuracy. This assumption is increasingly untenable due to the economic and physical limitations imposed by aggressive device scaling, which introduces heightened variability, noise susceptibility, and reliability issues~\cite{barkam2023reliable}.

\begin{figure}[t!]
    \centering
    \vspace{-2mm}
    \includegraphics[width=0.9\linewidth]{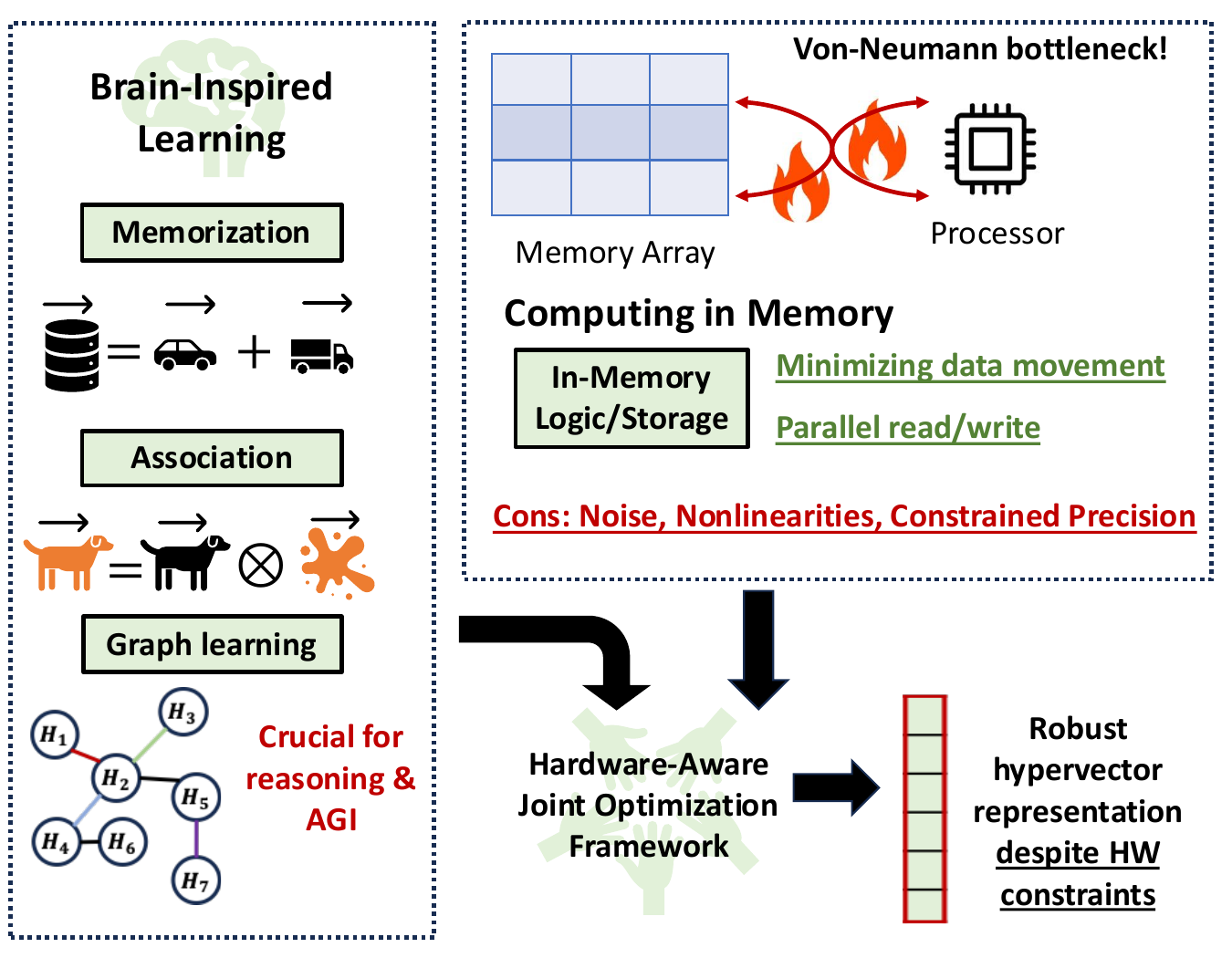}
    \vspace{-2mm}
    \caption{High-level overview of how our optimization framework bridges brain-inspired hyperdimensional computing and next-generation computing-in-memory architectures. Our optimization framework ensures that core hyperdimensional operations (binding, bundling, and associative search) remain robust despite the nonlinearity inherent to emerging semiconductor devices. This approach supports both classification and graph-based reasoning, laying the foundation for energy-efficient AI reasoning systems.}
    \label{fig:encoding}
    \vspace{-6mm}
\end{figure}
Recent developments, particularly in Computing-in-Memory (CIM), aim to address these hardware limitations by reducing data movement bottlenecks inherent to traditional von Neumann architectures. However, these emerging CIM platforms often exhibit considerable noise, variability, and reliability challenges, rendering traditional ML techniques incompatible due to their sensitivity to hardware-induced nonlinearities and errors~\cite{hamza_tcas_hdgim}. Prior research has proposed solutions that mitigate these challenges but rely on ideal, pre-encoded hypervectors tailored explicitly for specific environments, thus sacrificing practical applicability and limiting online learning capabilities.

Hyperdimensional Computing (HDC), a brain-inspired computational paradigm, offers a robust and flexible alternative uniquely positioned to address these critical issues \cite{kanerva2009hyperdimensional}. By representing data in high-dimensional vector spaces, HDC inherently tolerates hardware imperfections and nonlinearities, facilitating effective operation even on noisy and resource-constrained edge devices \cite{rahimi2016robust}. HDC's transparent and interpretable decision-making~\cite{bayesian_hamza} further distinguishes it from traditional deep neural networks that require structure in order to achieve similar interpretable transformations~\cite{shewmake2023visual, locatello2020object}.

GrapHD~\cite{poduval2022graph} and RelHD~\cite{kang2022relhd} are the extension of HDC into the domain of graph learning by representing nodes, edges, and their relational structures with high-dimensional vectors (hypervectors). At its core, GrapHD encodes relationships using variable binding operations to preserve semantic distinctions. This mechanism enables interpretable reasoning on graph-structured data because the binding and unbinding processes are explicitly defined, allowing for transparent inspection of how different node and edge attributes combine. HDC-based graph approaches stand at the cutting edge of explainable AI research and emerging AGI efforts, offering robust symbolic-like reasoning and enhanced interpretability.

Building upon these hyperdimensional graph-based methods, we propose a hardware-aware optimization framework that addresses the practical limitations imposed by future semiconductor technologies—namely, noise, nonlinearities, and constrained precision in physical hardware. Rather than assuming ideal pre-encoded hypervectors, our approach explicitly formulates the encoding stage as an optimization problem. We minimize the Frobenius norm between a target kernel (which encodes desired similarity or binding relationships among hypervectors) and a hardware-constrained kernel (which captures the actual relationships once nonlinearities are considered). This formulation holds clear advantages for HDC classification tasks, where approximate kernel shapes suffice to maintain accuracy. It becomes exponentially more critical in graph-centric HDC models. In these cases, even minor deviations in the binding and associative operations, core to hyperdimensional reasoning, can disrupt the delicate relational structures that enable interpretability and decidability. By introducing a novel joint optimization procedure, our framework ensures that hypervectors remain robust against hardware nonidealities, preserving the reliability of the binding mechanism. Consequently, we equip HDC-based graph approaches with a stable end-to-end pipeline that not only strengthens their ability to learn and reason but also remains viable on next-generation, ultra-low-power devices—thereby advancing robust AI hardware co-design and accelerating the adoption of reasoning methodologies.

\section{Related Work}

\subsection{Hyperdimensional Computing (HDC)}
Hyperdimensional computing is a brain-inspired paradigm that represents data with long random vectors, called \textit{hypervectors}, in a high-dimensional space. Core operations in HDC are defined as simple vector manipulations that mimic cognitive tasks such as memory and reasoning. These mathematically simple operations facilitate low-complexity training and inference, yet HDC has demonstrated strong performance in domains such as speech recognition, vision, and reinforcement learning~\cite{imani2022neural, ni2024efficient, ni2024heal} with a fraction of the compute and energy utilized by deep neural networks. Moreover, due to the high dimensionality and quasi-orthogonal properties of hypervectors, HDC is inherently robust to noise and low-precision errors. This property makes HDC especially suitable for implementation on emerging hardware, including non-volatile memories, where computational errors and analog variability are prevalent.

Early adaptations of HDC to graph data such as GrapHD features a cognitive memory for graph reconstruction, matching, and shortest-path queries~\cite{Poduval2022GrapHD}. Subsequent methods like HyperNode and HDGL focus on node-level tasks, achieving competitive accuracy with graph neural networks while offering significant computational speedups~\cite{li2023hypernode, Dalvi2024HDGL}.

The work in~\cite{Han2024CiliaGraph} introduced symmetric neighborhood aggregation and node attributes to boost HDC expressivity for graph data, attaining performance comparable to modern GNNs. HyperGRAF optimized graph reasoning for FPGA implementation, surpassing GPU-based designs in both speed and energy efficiency~\cite{Chen2023HyperGRAF}. Collectively, these developments highlight the efficiency, accuracy, and interpretability of graph-based HDC, making it a robust choice for resource-constrained and noisy environments.

\subsection{FeFET-Based In-Memory Hyperdimensional Computing}
Traditional von Neumann architectures face the \textit{memory wall}, where processor-memory data transfers dominate latency and energy \cite{PIM_GLS_li2019overview}. CIM mitigates this bottleneck by integrating processing units directly within memory arrays. Two main categories have emerged: (1) \textbf{Analog CIM}, which uses resistive memory devices like RRAM \cite{RRAM_review} and PCM \cite{PCM_Review} to execute parallel matrix-vector multiplications using electrical properties (Ohm's law and Kirchhoff's law) of the devices, but often requires analog-to-digital conversion \cite{yu2018neuro, shafiee2016isaac}; (2) \textbf{Digital CIM}, which performs Boolean logic operations using resistive switching of memory bitcells to avoid data conversion overhead, at the cost of limited instruction sets and additional cycles for multi-bit operations \cite{hamdioui2015memristor}. Recent hybrid designs combine analog computing for core tasks with digital control for thresholding and binding, achieving high parallelism and robustness to noise \cite{rashed2021hybrid}.

FeFETs, which integrate an HfO\(_2\)-based ferroelectric layer into the gate stack, have emerged as a promising candidate device for CIM architectures, due to their non-volatile nature, CMOS compatibility, ability to support both storage and logic operations \cite{huang2023fefet, marchand2021fefet} and much higher endurance, retention, low drift compared to other non-volatile memories. Polarization-dependent threshold voltage shifts allow each FeFET to function as a memory element. While recent studies have harnessed FeFETs for in-memory logic and CAMs \cite{cai2024scalable}, variability in the ferroelectric layer remains challenging, prompting circuit-level strategies like adding a selector transistor (1T1R) to mitigate threshold voltage fluctuations \cite{parmar2023demonstration}.  

HDC’s inherent noise tolerance makes it well-suited for deployment on in-memory FeFET arrays which are intrinsically noisy. Early in-memory HDC was demonstrated on resistive memory (e.g., PCM) \cite{karunaratne2020memory}, while Huang \textit{et al.} \cite{huang2023fefet} employed FeFET crossbars for binding, permutation, and bundling, achieving a 5.04\(\times\) energy efficiency gain over memristors. Although analog in-memory solutions using RRAM or PCM have shown promise of energy-efficient HDC, FeFET-based architectures promise higher speed, energy efficiency, better scalability and integration potential. Table \ref{tbl:speedup} highlights the recent works of FeFET-Based HDC accelerators and the expected speedup and energy efficiency versus NVIDIA GPUs. The exponential speedup and energy efficiency of FeFET-based HDC accelerators make them a compelling path forward, enabling low-power learning and reasoning. This efficiency makes large-scale HDC deployments feasible and provides a scalable and efficient architecture for advanced reasoning systems in the future.

\begin{table}[h]
\centering
\vspace{-4mm}
\caption{HDC speedup}
\label{tbl:speedup}
\vspace{-3mm}
\begin{tabular}{cccc} 
\toprule
\textbf{Ref.} & \textbf{Speedup} & \textbf{Energy efficiency} \\
\midrule
RelHD vs 1080Ti~\cite{kang2022relhd} & 33x & 59287x \\
FeReX vs 3090Ti~\cite{xu2024ferex} & 250x & 18000x \\
MIMHD vs 1080~\cite{kazemi2021mimhd} & 56x & 314x \\
\bottomrule
\end{tabular}
\vspace{-4mm}
\end{table}
\section{Hyperdimensional Computing Encoding}
Fourier Holographic Reduced Representations (FHRR) are a class of hyperdimensional embeddings in which high-dimensional vectors (hypervectors) are defined in the complex domain, typically on the unit circle. Each component \(v_i\) of an FHRR hypervector \(\mathbf{v}\) is represented as $v_i = e^{j \theta_i}, \quad \theta_i \sim \text{Uniform}(0, 2\pi)
$, where \(j\) is the imaginary unit, and \(\theta_i\) is the phase. Since the magnitude of \(v_i\) is fixed at 1, the phase \(\theta_i\) encodes the primary information.

HDC relies on two fundamental operations: \emph{bundling} (superposition) and \emph{binding} (association). \textbf{Bundling (Superposition):} Hypervectors are added component-wise, $\mathbf{v}_\text{bundle} = \sum_k \mathbf{v}_k,$ followed by normalization to preserve dimensional magnitudes. \textbf{Binding (Association):} Hypervectors are multiplied component-wise, $\mathbf{v}_\text{bind} = \prod_k \mathbf{v}_k = e^{l(j \sum_k \theta_k)}$. Binding thus translates to adding the phases of the vectors being multiplied. The inverse of a hypervector \(\mathbf{v}\) is simply its complex conjugate when \(\|\mathbf{v}\|=1\), enabling straightforward unbinding via \(\mathbf{v}^{-1} = \overline{\mathbf{v}}\).

FHRR’s phase-based, continuous encoding facilitates smooth interpolation and rotation-based manipulations while maintaining robustness to noise via the unit-magnitude constraint. Binding and unbinding map directly to phase addition and conjugation, respectively, enabling efficient compositional reasoning for tasks like sequence modeling, logic-based inference, and symbolic manipulation. Graphs are encoded by binding node and edge hypervectors, and can be “unrolled” to retrieve subgraphs via conjugation. Compared with binary or real-valued hyperdimensional approaches, FHRR streamlines inverse operations and handles continuous attributes more effectively, making it particularly suitable for tasks that demand both symbolic and numerical flexibility \cite{plate2003holographic,kanerva2009hyperdimensional}.

\subsection{Graph-Based HDC}\label{sec:graph_hdc}

Graph-based HDC extends the encoding to structured data by encoding nodes into hypervectors and utilizing the binding operation to associate relations and edges between nodes. This enables an efficient graph memorization and allows efficient HDC operations to reason over such graph structures. A visualization of HDC-based graph and edge generations can be found in figure~\ref{fig:encoding}. We specifically study the capabilities of our joint optimization framework applied to two popular graph-based HDC models, GrapHD~\cite{poduval2022graph} and RelHD~\cite{kang2022relhd}.

\subsubsection{\textbf{GrapHD}}\label{subsub:graphd}
GrapHD is a HDC-based graph model that maps graph-based information into high-dimensional representations and enables cognitive computing over a final graph hypervector. Unlike traditional GNNs that rely on expensive iterative message passing, GrapHD encodes the structural information directly into hypervectors, allowing for efficient graph reasoning and retrieval. GrapHD specifically utilizes the binding operation to associate a node with its connections. In brief, this is done with the following equation:
\begin{equation}
\label{eq:graphd}
G = \frac{1}{2}(H_{1} * M_{1} + H_{2} * M_{2} + \dots + H_{n} * M_{n})
\end{equation}
where $G$ is the graph representation, $H_i$ is the representation of the $ith$ node, and $M_i$ is the memory connection hypervector that contains the information of all nodes connected to the $ith$ node. $M_i$ is simply the bundling of all the node hypervectors that have an edge with $H_i$. 
A node's memory can be reconstructed by binding the node hypervector with the graph hypervector:
\begin{equation}
\label{eq:grapHD_node_reconstruction}
H_i * G = H_i * \frac{1}{2}\big(\sum_{j=1}^nH_j * M_j\big)
\end{equation}
where the property of self-binding (i.e. $H_i * H_i \simeq 1$) and $H_i$ and $H_j$ are ideal (approximately orthogonal), equation~\ref{eq:grapHD_node_reconstruction} can be simplified to:
\begin{equation}
\label{eq:graphd_simplified}
H_i * G = M_i + \epsilon
\end{equation}
where $\epsilon$ represents some accumulated noise from the approximate orthogonality of the hypervectors, but if all the representations are noisy, $H_i$ fails to recover properly recover $M_i$. Thus, robust representations are crucial for GrapHD and its ability to retrieve the relations between nodes.

\subsubsection{\textbf{RelHD}}\label{subsub:relHD}
While GrapHD efficiently encodes and associates local neighboring relations with nodes, often times multi-hop relations are essential in order to preserve the structure of more complex graph structures. Thus, RelHD introduces a different HDC graph encoding scheme by introducing hierarchical encoding to explicitly differentiate between 1-hop and 2-hop neighbors.

RelHD extends the graph representation such that a relation hypervector $I_k$ is formed by aggregating a node hypervector, $M_k$, with its 1-hop neighbor hypervector, \( M_k^1 \), and its 2-hop neighbor hypervector, \( M_k^2 \).
\( M_k^1 \) can be computed the same way as GrapHD where it is just the bundling of all neighbors directly connected to the node hypervector $H_k$. Iteratively searching for the 2-hop neighbors can be computationally demanding and thus \( M_k^1 \) is reused to build the 2-hop neighbor representation \( M_k^2 \). Specifically, \( M_k^2 \) is computed by the following:

\begin{equation}
    M_k^2 = \sum_{j \in B} M_j^1.
\end{equation}
where B is the set of all direct neighbors of $H_k$ instead of \( M_k^1 \) to save computation.

The final relation hypervector (\( I_k \)) is computed using the binding operation:
\begin{equation}
    I_k = H_k * \omega_0 + M_k^1 * \omega_1 + M_k^2 * \omega_2.
\end{equation}
where $\omega_0, \omega_1, \omega_2$ are orthogonal bipolar hypervectors (i.e. every element in $\omega$ is randomly sampled to be either -1 or 1). This is used to distinguish the information in $I_k$ such that $H_k$, $M_k^1$, or $M_k^2$ can be retrieved by binding $I_k$ with the corresponding $\omega$. RelHD has specifically shown strong results in terms of accuracy and efficiency for node classification on graph structure datasets. 

\section{Framework}

\begin{figure*}[t!]
    \centering
    \vspace{-4mm}
    \includegraphics[width=0.8\linewidth]{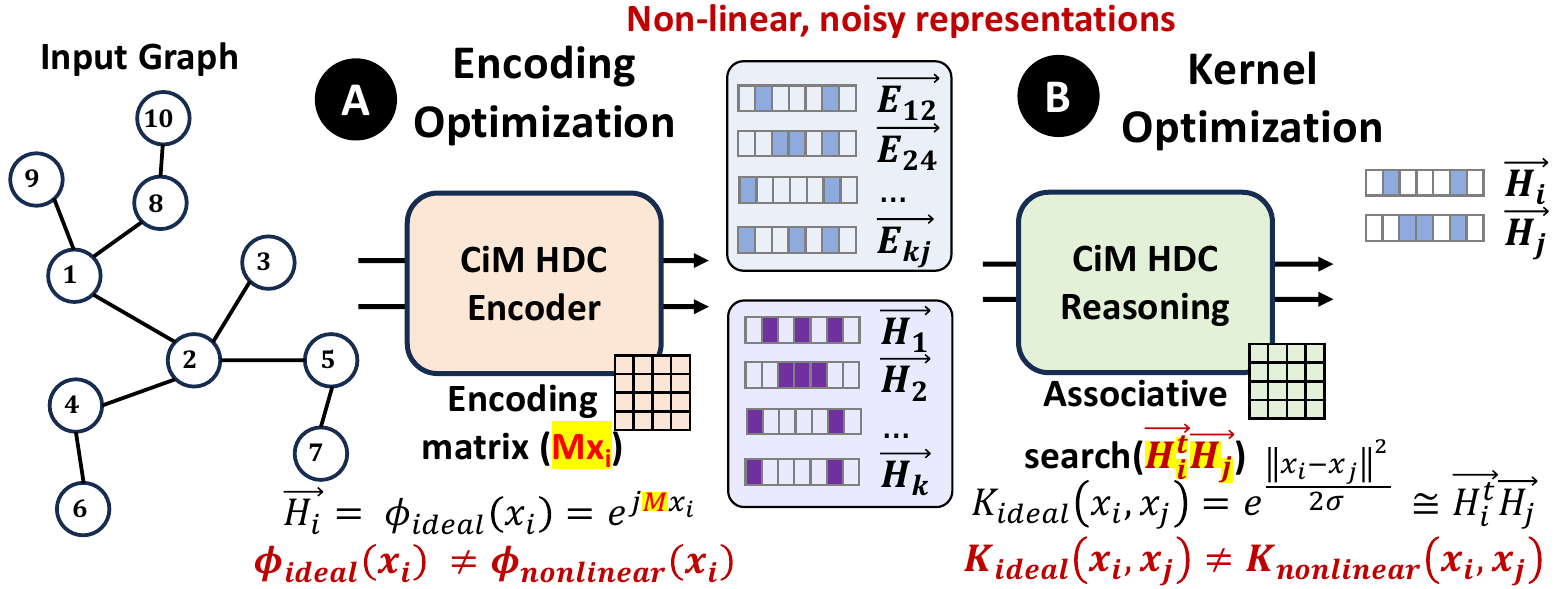}
    \vspace{-3mm}
    \caption{Hardware-aware HDC using our joint-optimization framework. The system optimizes encoding under nonlinear constraints from CiM hardware, ensuring robust associative search and reasoning through kernel optimization.}

    \label{fig:framework}
    \vspace{-2mm}
\end{figure*}

\subsection{Nonlinear Similarity in CIM Hardware}
Hyperdimensional Computing traditionally assumes that similarity computations between hypervectors can be measured with idealized metrics like cosine similarity~\cite{kanerva2009hyperdimensional}. However, CIM architectures often implement these comparison operations with in-memory analog operations that deviate from the ideal linear behavior. For example, many CIM designs approximate cosine distance using binary dot-product circuits or content-addressable memories~\cite{liu2020hardware}. These non-ideal implementations introduce \emph{nonlinearities and noise} (e.g., partial analog summation across subarrays can saturate or incur significant noise) causing the hardware-measured similarity to differ from the true cosine similarity. Our framework addresses this gap by explicitly accounting for the hardware’s \emph{effective similarity function} \( S_{\text{hw}} \) that captures distortions (such as quantization, saturation, and device noise), ensuring that the encoding and search operation are performed in hardware-aware manner, considering the realistic constraints of CIM hardware rather than an idealized metric.

\subsection{Joint Optimization of Encoding and Similarity Search}
Rather than designing the hypervector encoding and similarity search independently, we propose a \emph{joint optimization} strategy that treats the encoding module and the associative search module as a unified learnable system. In standard HDC pipelines, the encoding module is fixed after initialization and the search module simply compares hypervectors using a set distance metric~\cite{kanerva2009hyperdimensional}. In contrast, our approach simultaneously optimizes the encoding function \( \phi_\theta \) and the similarity retrieval process to compensate for hardware-induced deviations. The key idea is to train the hypervector representations such that when the CIM hardware computes their similarity, the correct results are obtained despite nonlinearities. Concretely, we include the associative search behavior in the training loop: during optimization, the output of the hardware-modeled similarity search (the retrieval of nearest hypervectors or class labels) feeds into the loss function. This end-to-end optimization aligns the encoding of data and the similarity comparison mechanism, mitigating errors that would arise from non-ideal hardware operations. By co-optimizing both stages, the system learns embeddings that are maximally robust to hardware effects, effectively \emph{calibrating the similarity search} to the encoded hypervectors and vice versa.

We formalize the above ideas with a joint objective that explicitly models non-ideal encoding and similarity computations. Let $\phi_\theta: \mathcal{X} \to \mathbb{R}^D$
be the learnable encoding function mapping an input to a \(D\)-dimensional hypervector. To capture hardware non-idealities, we define a hardware distortion function \( f: \mathbb{R}^D \to \mathbb{R}^D \) that transforms an ideal hypervector into its stored analog within the CIM hardware (accounting for quantization errors, nonlinear transfer functions, etc.). Given two inputs \( x_i \) and \( x_j \) with encoded vectors \( v_i = \phi_\theta(x_i) \) and \( v_j = \phi_\theta(x_j) \), the \emph{hardware-perceived similarity} is modeled as:
\begin{equation}
\label{eq:hardwaresim}
S_{\text{hw}}(x_i, x_j) = \Psi\big( f(v_i),\, f(v_j) \big)
\end{equation}
where \( \Psi \) denotes the similarity operation as executed in hardware (for instance, an analog dot product or Hamming distance circuit). This formulation extends the ideal cosine similarity to a \emph{hardware-aware similarity} \( S_{\text{hw}} \) that includes nonlinear effects.
\begin{figure*}[h]
    \centering
    \begin{subfigure}[t]{0.49\textwidth}
        \centering
        \includegraphics[width=\linewidth]{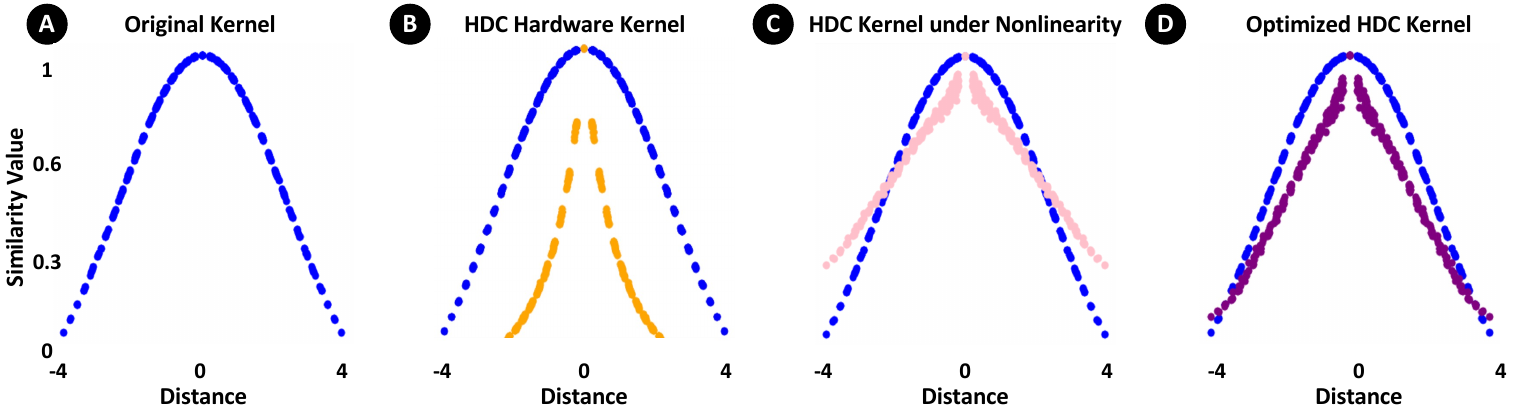}
        \caption{HDC kernel shape under nonlinear associative search}
    \end{subfigure}%
    \hfill
    \begin{subfigure}[t]{0.49\textwidth}
        \centering
        \includegraphics[width=\linewidth]{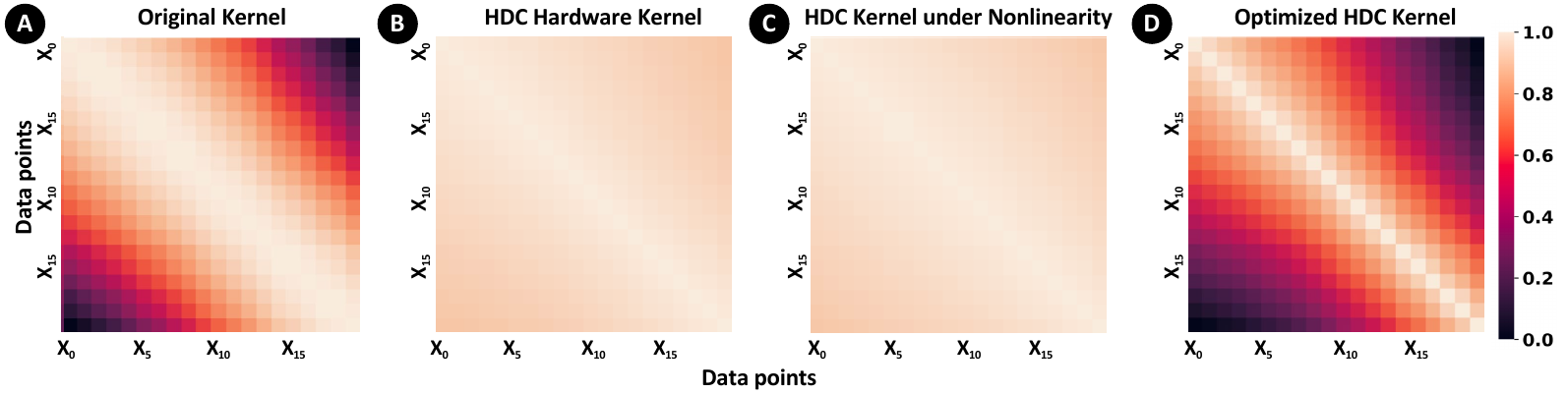}
        \caption{HDC kernel heatmaps under both nonlinear encoding and associative search}
    \end{subfigure}
    \vspace{-4mm}
    \caption{(a) are the kernel shapes under only nonlinear associative search while (b) are the kernel heatmaps under both nonlinear encoding and associative search process.
    (A) represents the original desired kernel (B) shows HDC kernel approximation (C) is the similarity-optimized HDC approximation (D) is our proposed joint optimization kernel approximation.}
    \label{fig:kernel_approximation}
    \Description{Two side-by-side visualizations: one showing kernel shapes under nonlinear associative search, and one showing heatmaps under nonlinear associative search and encoding.}
    \vspace{-4mm}
\end{figure*}

We then introduce a target similarity matrix \( \mathbf{S}^* \) to encode the desired similarities between all pairs of encoded items (e.g., \( \mathbf{S}^*_{ij}=1 \) if \( x_i \) and \( x_j \) should be similar, and 0 if dissimilar, based on ground truth or task requirements). Our goal is to make the hardware-computed similarities align with \( \mathbf{S}^* \). To achieve this, we define a loss function that penalizes differences between actual hardware similarities and the target values:
\begin{equation}
\label{eq:simloss}
\mathcal{L}_{\text{sim}}(\theta) = \sum_{i,j} \Big( S_{\text{hw}}(x_i, x_j) - \mathbf{S}^*_{ij} \Big)^2.
\end{equation}
By minimizing \( \mathcal{L}_{\text{sim}} \), the encoding parameters \( \theta \) are directly adjusted so that \( f(\phi_\theta(x)) \) produces hypervectors whose CIM-level similarity \( \Psi \) matches the idealized relationships in \( \mathbf{S}^* \).

In practice, \( \mathcal{L}_{\text{sim}} \) is combined with task-specific loss terms (e.g., a classification loss ensuring correct label retrieval) to form a joint objective:
\begin{equation}
\label{eq:practiceloss}
\min_{\theta} \; \mathcal{L}_{\text{task}}(\theta) + \alpha\, \mathcal{L}_{\text{sim}}(\theta) + \beta\, \mathcal{R}(\theta),
\end{equation}
where \( \mathcal{L}_{\text{task}} \) is a task loss (e.g., classification error), \( \mathcal{R}(\theta) \) is a regularization term (discussed next), and \( \alpha,\beta \) are hyperparameters balancing the different objectives.

Our framework achieves robustness across diverse CIM architectures and noise profiles by modeling each hardware’s nonlinearity in the similarity function (via \(f\) and \(\Psi\)) and optimizing over an \emph{ensemble} of hardware conditions. During training, noise is sampled or hardware parameters are varied to minimize the expected loss \(E_{\eta \sim \mathcal{N}}[\mathcal{L}_{\text{sim}}(\theta; \eta)]\), where \(\eta\) represents random noise or process variation. This teaches the encoder to produce hypervectors resilient to perturbations. To enable this, we learn a flexible encoding function \(\phi_\theta\) (e.g., a small neural network), rather than relying on fixed or randomly initialized base hypervectors. By parameterizing \(\phi_\theta\), the system can adapt each hypervector component to maximize hardware-aware similarity outcomes, thereby yielding robust embeddings even under hardware distortions. Finally, we introduce a regularization term \(\mathcal{R}(\theta)\) that enforces well-behaved embeddings with approximate normalization and sparsity, preventing extreme values or saturation while preserving the distributed, nearly orthogonal properties of hyperdimensional representations.
\section{Results}

Graph-based reasoning is increasingly recognized as a foundational building block for advanced cognitive tasks and a critical stepping stone toward Artificial General Intelligence. Structured data representations enable models to capture relational knowledge, facilitate complex inferencing, and promote interpretability. Given these requirements, we include graph-focused benchmarks in our evaluation to demonstrate how our framework supports robust and scalable graph reasoning, laying the groundwork for more generalizable and cognitively rich AI systems. We rigorously evaluate our framework on four primary benchmark tasks to demonstrate its effectiveness and generality: 
\begin{itemize}[leftmargin=*]
    \item \textbf{Kernel Approximation:} We assess how well the framework approximates target kernels under varying nonlinearities. 
    \item \textbf{HDC Classification:} We compare our learned encoders against standard HDC baselines on common datasets, measuring classification accuracy and resilience to hardware noise.
    \item \textbf{Graph Reconstruction:} We verify the robustness of relation embeddings by examining how accurately our model reconstructs graph structures and similarities.
    \item \textbf{Graph Node Prediction:} We further test knowledge representation by evaluating node-level predictions on graph benchmarks.
\end{itemize}

\begin{figure*}[t!]
    \centering
    \includegraphics[width=0.7\linewidth]{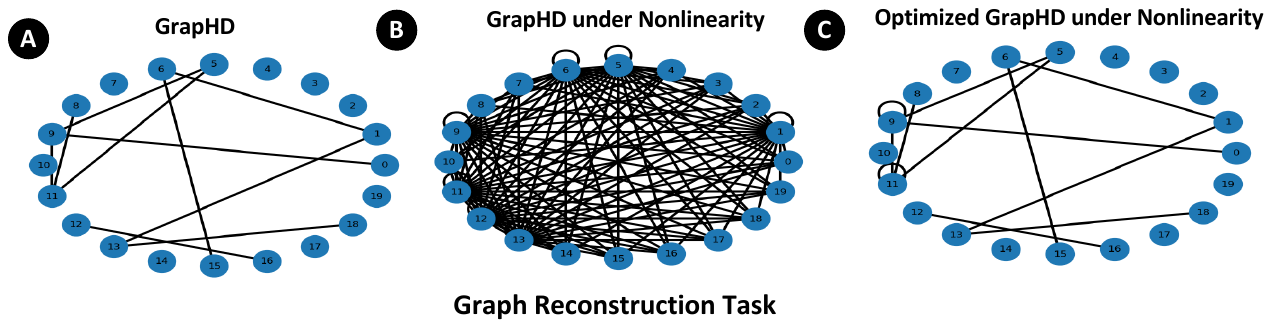}
    \vspace{-3mm}
    \caption{GrapHD-based Graph reconstruction under hardware nonlinearity and noise. We see that a (B) GrapHD under hardware-induced nonlinearity fails to reconstruct the true graph as it is not able to classify edges between nodes properly. (C) Our optimization allows GrapHD to work even under nonlinearity.}
    \vspace{-3mm}
    \label{fig:GrapHD Graph Reconstruction}
    \vspace{2mm}
\end{figure*}
A key focus is the benefit of our \emph{joint optimization} scheme, where we apply the framework to existing HDC-based models under hardware uncertainty. This joint approach consistently improves performance for both image classification and reasoning on graph-structured data. All experiments utilize cosine similarity as the standard HDC comparison metric unless specified otherwise, while the nonlinearity \(\Psi\) is instantiated as an exponential, logarithmic, or hyperbolic tangent noise function. In practical CIM systems, \(\Psi\) can be substituted with any nonlinearity specific to the target hardware device.

We implement our framework in \texttt{Python} using the \texttt{PyTorch} library and repeat each experiment 10 times, reporting mean accuracy. These settings ensure reproducibility, demonstrating that our technique is readily deployable across diverse CIM platforms and device characteristics.

\subsection{Kernel Approximation}\label{subsection:kernel_approximation}
The kernel approximation study evaluates the impact of adding noise to essential HDC operations with nonlinearity. We utilize a synthetic dataset, comprised of 20 data points with each sample having 30 features. The first data point is sampled from a uniform distribution $U(-1, 1)^{30}$ while each subsequent point is generated by adding small perturbations sampled from $U(0, 0.1)^{30}$. Figure~\ref{fig:kernel_approximation} illustrates how different HDC approaches influence the kernel approximations. Original kernel (A) represents the desired Radial Basis Function (RBF) kernel:
\begin{equation}
\label{eq:rbf}
    K(x, y) = exp(-\frac{||x - y||^2}{2\sigma^2})
\end{equation}
where $||x - y||^2$ is considered the squared Euclidean distance between the two feature vectors and $\sigma$ controls the smoothness and spread of the kernel function. For all experiments, $\sigma$ is set to be $\frac{1}{n}$ where $n$ is the number of data points. As expected, we see that as the distance between the data points increases, the similarity values begin to decrease. 

HDC Hardware Kernel (B) is when HDC uses the hardware-induced similarity function, \(\Psi\), instead of cosine similarity for the associative search module. In figure~\ref{fig:kernel_approximation}(a), we see that while it is able to recover that the same data points are very similar (i.e. similarity = 1) it severely undervalues the similarities very quickly when the distance between the data points increases and the kernel shape is significantly distorted as a result.

HDC Kernel under Nonlinearity (C) optimizes the HDC associative search module by minimizing the loss between the desired target RBF kernel values and the HDC similarity values under \(\Psi\). The loss equation can be thought of using equation~\ref{eq:simloss}, but replacing $S_{\text{hw}}(x_i, x_j)$ with the RBF kernel values. In essence, we simply optimize (B) to approximate (A). The optimization helps approximate the shape of the desired kernel but loses the parabolic shape, resulting in a more linear approximation.

Optimized HDC Kernel (D): Our proposed method optimizes the similarity representation through the our joint optimization approach detailed in our framework section and equation~\ref{eq:simloss}. We minimize the Frobenius norm between the HDC hardware kernel (B) and the hardware computed similarity matrix, and thus considers the target similarity and encoding are under similar hardware constraints. This approach restores much of the original structure despite the constraints imposed by nonlinearity. Specifically, (D) is much better than both (B) and (C) at approximating the tail-ends of the kernel shapes. This means that (D) is much more efficient at maintaining the dissimilarity between farther data points. We attribute this to the fact that (D) also utilizes a regularization term and also has more flexibility due to the joint optimization approach.

In figure~\ref{fig:kernel_approximation}(b), we consider a situation where one wants to deploy an end-to-end framework on emerging hardware. In this scenario, hardware-induced nonlinearities are applied at both stages of the HDC pipeline, encoding and associative search. Figure~\ref{fig:kernel_approximation}(b) again highlights the desired original kernel (A), the HDC hardware kernel (B), HDC kernel under nonlinearity (C), and our proposed optimized HDC kernel (D). When the encoding and associative search modules are changed to \(f(\cdot)\) and \(\Psi\), we see that both (B) and (C) fail to approximate the desired original kernel. We attribute this to the fact that the encoding process also contains noise from non-ideal hardware operations and the errors exponentially increases as the distance between the becomes larger. As mentioned earlier, our joint optimization approach is specifically better at maintaining the dissimilarity between farther points and is able to consider calibrating both the encoding and associative search together. 

\begin{figure}[t!]
    \centering
    \vspace{-4mm}
    \includegraphics[width=\linewidth]{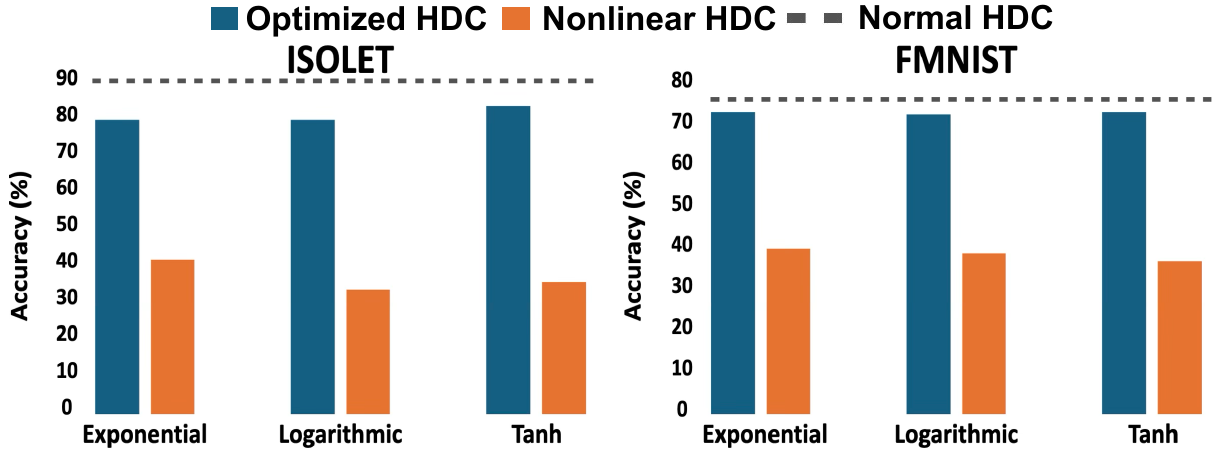}
    \vspace{-3mm}
    \caption{Our proposed HDC optimization for HDC classification compared to naive HDC implementation on noisy Hardware. The dotted represents QuantHD's normal performance on Isolet and FMNIST datsets.}
    \label{fig:quanthd_learning}
    \vspace{-5mm}
\end{figure}

\subsection{HDC Classification}\label{subsection:HDC_classification}
HDC has shown strong results for edge-friendly classification due to its inherent robustness to noise and fast single-pass learning through lightweight vector operations~\cite{hernandez2021onlinehd}. QuantHD~\cite{imani2019quanthd} further optimizes HDC by using low-cost quantized models while still maintaining accuracy, making it ideal for hardware-constrained or low power settings. We test our framework on popular HDC-based classification benchmarks such as speech recognition, ISOLET~\cite{isolet_54}, and image classification, FMNIST~\cite{xiao2017fashion}. We consider end-to-end learning on noisy hardware and showcase the results in figure~\ref{fig:quanthd_learning}.

We utilize a dimensionality of 512 for QuantHD and quantize using the sign function, assigning all positive and negative elements to 1 and 0 bits. QuantHD also utilizes the hamming distance instead of cosine similarity for its associative search module. Our proposed optimization, when applied to QuantHD, shows a significant increase over the naive QuantHD implementation when replacing the associative search with different \(\Psi\). Our optimized QuantHD is able to achieve an impressive 84\% on ISOLET and 73\% on FMNIST even with a nonlinear Tanh function applied to the encoding and associative search modules. A naive implementation of QuantHD is still able to achieve almost 37\% and 36\% accuracy on ISOLET and FMNIST under the same nonlinear conditions. We attribute this to HDC is being robust to noise and HDC-based classification only relying on bundling and similarity operations. Because of this, HDC can learn regardless of the noise added from the nonlinear encoding module, and the drop in performance is mainly due to the noise of the associative search module. Therefore, in the next section, we consider HDC applications where the binding operation is utilized and precise high-dimensional representations are required.

\begin{figure}[t!]
    \centering
    \vspace{-1mm}
    \includegraphics[width=\linewidth]{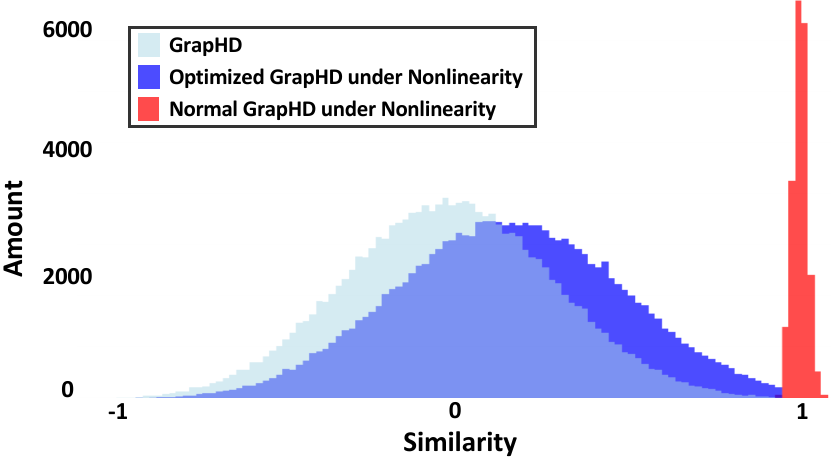}
    \vspace{-5mm}
    \caption{Similarity distributions between nodes for (A) GrapHD (B) GrapHD under Nonlinearity (C) Optimized GrapHD under Nonlinearity. Our optimization gives GrapHD a similar distribution even under hardware constraints while (B) has an extremely narrow distribution centered around 1.}
    \label{fig:graph_distribution}
    \vspace{-7mm}
    \Description{A visualization of GrapHD's similarity distribution under normal and nonlinearity}
\end{figure}

\subsection{Graph Reconstruction}\label{subsection:graph_reconstruction}
In this section, we explore our optimization framework to enable graph-based tasks under nonlinearity. We study GrapHD's ability to accurately represent and reconstruct graph structures when the encoding and associative search models are noisy.

In figure~\ref{fig:GrapHD Graph Reconstruction}, we test the graph reconstruction abilities of GrapHD with and without our optimization. We randomly generated 20 node graph with 10 edges as our target graph and proceed to test grapHD's reconstruction capabilities in various situations. GrapHD reconstructs the true graph perfectly under normal circumstance in (A). However in (B), GrapHD under nonlinearity fails to recover the true graph and misclassified all the edge connections. Yet in (C), our optimization ensures that the grapHD representations are robust to the nonlinearity applied at the encoding and associative search modules. The reasoning behind the poor results under nonlinearity is shown in figure~\ref{fig:graph_distribution}. In figure~\ref{fig:graph_distribution}, we check the similarity distribution between all the node hypervectors and notice that the distribution is extremely narrow and centered around 1 for (B) while the optimized GrapHD (C) has a similar distribution to GrapHD under normal circumstances. Specifically, what is happening is that the noisy representations are pushing the similarity between all node hypervectors too high such that equation~\ref{eq:grapHD_node_reconstruction} turns into:
\begin{equation}
\label{eq:similarity_reconstruction}
H_i * G = \frac{1}{2}\big(\sum_{j=1}^ns_{ij} * M_j\big)
\end{equation}
where $s_{ij}$ is the similarity between $H_i$, and $H_j$. If $s_{ij} \simeq 1$, which is the case under the nonlinearity \(f(\cdot)\) and \(\Psi\), this becomes the sum of all memory hypervectors, resulting in GrapHD under nonlinearity (B) where nodes are connected to every other node during the graph reconstruction task. On the other hand, our optimization framework ensures GrapHD still creates accurate representations under nonlinearity and noise which can be utilized for any downstream GrapHD reasoning tasks such as memory reconstruction, information retrieval, graph matching, and shortest path.

\begin{figure}[t!]
    \centering
    \vspace{-4mm}
    \includegraphics[width=0.8\linewidth]{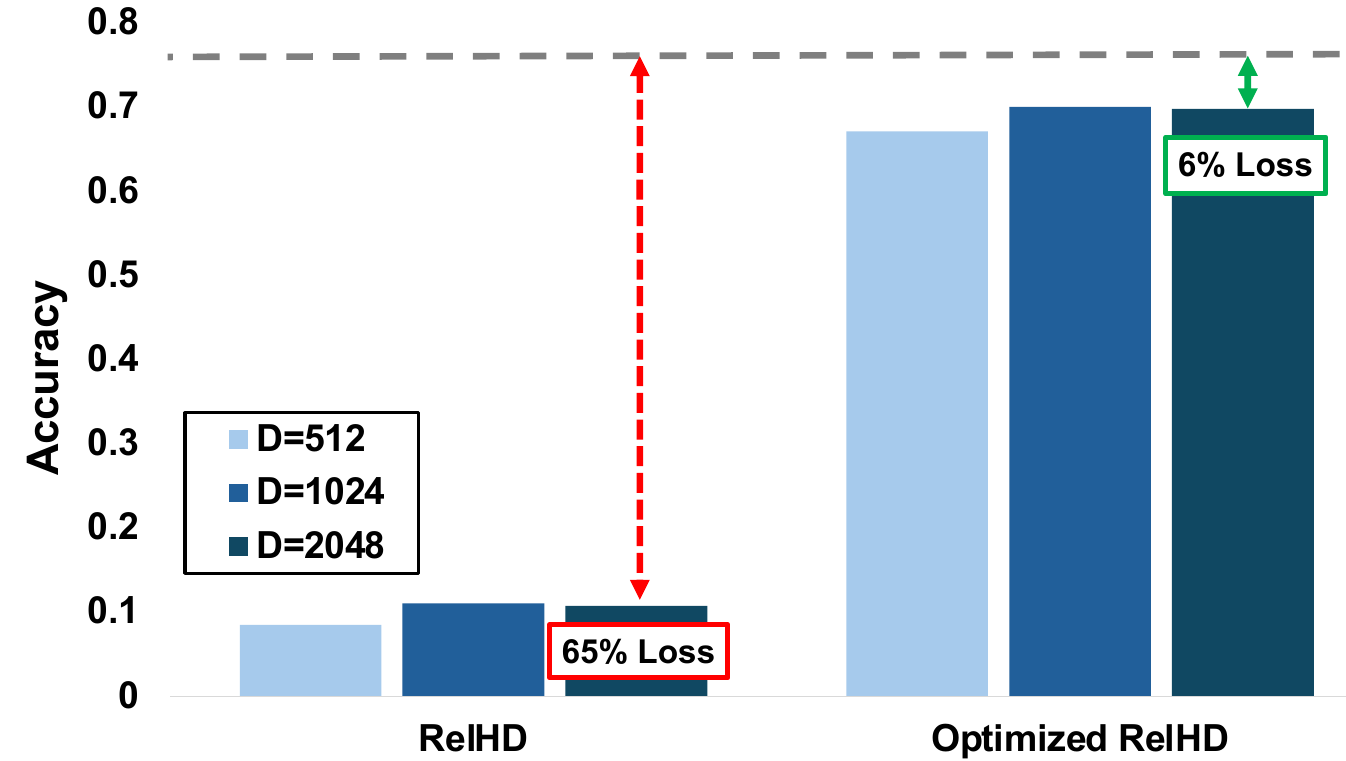}
    \vspace{-3mm}
    \caption{Our proposed optimization for RelHD compared to naive RelHD implementation on noisy hardware. The dotted line represents the RelHD performance under normal environments.}
    \label{fig:relhd}
    \vspace{-5mm}
\end{figure}

\subsection{Graph Reasoning Tasks}\label{subsection:graph reasoning tasks}
Figure~\ref{fig:relhd} highlights our optimization results on the Cora dataset~\cite{McCallum2000Cora}, which is widely used for evaluating graph-based models. The Cora dataset consists of a citation network where nodes represent research papers and edges denote the citation relationships between them. Each node also has a feature vector associated from its text. Given its graph-based structure, the Cora dataset is well-suited for testing models that leverage relational reasoning, message passing, and neighborhood aggregation. RelHD focuses on node classification on the Cora dataset and utilizes HDC to learn the global and local relations between the nodes. When compared to a naive implementation of relHD under the hardware-induced nonlinearity, we see a significant performance benefit of our optimization framework. When compared to relHD's performance under normal environments (i.e. a Nvidia RTX 3060ti GPU), we see that our optimized RelHD only has 6\% quality loss while a naive implementation has a 65\% quality loss. Additionally, our results show that the dimensionality does not have a large effect on the performance as our optimized relHD achieves a 67\% and 70\% accuracy with $D=512$ and $D=2048$ respectively, making our optimized relHD suitable for low memory devices. Our optimization approach allows relHD to generate robust representations and relation embeddings under nonlinear environments, which is essential for downstream tasks such node classification and link prediction.

\section{Conclusion}
Implications of our findings for deploying learning algorithms on next-generation accelerators. Our results highlight the importance of incorporating hardware-induced noise into the learning process for robust deployment on CIM accelerators. By jointly optimizing encoding and similarity search under non-ideal conditions, we provide a framework that enables effective learning in noisy environments. The joint optimization framework demonstrates exponential improvements to both kernel approximation, learning on noisy hardware, and can be applied to any existing HDC-based models. Our optimization achieves an impressive 59\% improvement when applied to a graph-based HDC model under hardware-induced nonlinearity and 48\% improvement for HDC-based classification models. Additionally, our optimization fixes the relational binding problem for HDC graph models, allowing for downstream reasoning tasks on CiM accelerators. Future work will explore extending our approach to more complex reasoning tasks and further optimizing computational efficiency for large-scale applications.

\begin{acks}
    This work was supported in part by the DARPA Young Faculty Award, the National Science Foundation (NSF) under Grants \#2127780, \#2319198, \#2321840, \#2312517, and \#2235472, the Semiconductor Research Corporation (SRC), the Office of Naval Research through the Young Investigator Program Award, MURI grant, and the National Defense Science \& Engineering Graduate (NDSEG) Fellowship Program. Additionally, support was provided by the Air Force Office of Scientific Research under Award \#FA9550-22-1-0253 and Army Research Office Grant \#W911NF2410360. 
\end{acks}

\bibliography{bib.bib}
\bibliographystyle{ACM-Reference-Format}

\end{document}